\documentclass[prb, twocolumn, superscriptaddress, floatfix]{revtex4-2}
\usepackage[english]{babel}

\usepackage{amssymb}
\usepackage{amsmath}
\usepackage{amsfonts}
\usepackage{bm}
\usepackage{tabularray}
\usepackage[colorlinks, allcolors=blue]{hyperref}
\usepackage{graphicx}
\usepackage{xcolor}
\usepackage{tikz}
\tikzset{font=\footnotesize}
\usepackage{pgfplots}
\pgfplotsset{
    compat=1.18,
    layers/my layers/.define layer set={
        main,
        foreground
    }{},
    correlator/.style = {mark=*, only marks, mark size=1.5pt},
    fit/.style = {white!70!black, very thick, domain=1:80, no marks, samples=200, smooth, on layer=foreground},
}
\usepackage{array}
\usepackage{booktabs}
\usepackage[caption=false]{subfig}
\usepackage[shortlabels]{enumitem}
\usepackage{siunitx}[=v2]
\usepackage{physics}
\ExplSyntaxOn
\msg_redirect_name:nnn { siunitx } { physics-pkg } { none }
\ExplSyntaxOff

\newcommand{\floor}[1]{\left\lfloor #1 \right\rfloor} 

\newcommand{\FQ}{F_{\text{Q}}} 
\newcommand{\fQ}{f_{\text{Q}}} 

\renewcommand{\tilde}{\widetilde}

\definecolor{RoyalBlue}{HTML}{0071BC}
\definecolor{NavyBlue}{HTML}{006EB8}
\definecolor{VaBlue}{HTML}{023B88}
\definecolor{QuOrange}{HTML}{D87237}
\definecolor{Qurosso}{RGB}{198,62,62}
\definecolor{Naverde}{RGB}{47,128,0}
\definecolor{YellowGreen}{HTML}{98CC70}
\definecolor{SeaGreen}{HTML}{3FBC9D}
\definecolor{OliverGreen}{HTML}{3C8031}
\definecolor{PineGreen}{HTML}{008B72}
\definecolor{GGreen}{HTML}{00A64F}
\definecolor{LimeGreen}{HTML}{8DC73E}
\definecolor{GreenYellow}{HTML}{DFE674}
\definecolor{cadmiumgreen}{rgb}{0,0.42,0.24}
\definecolor{calpolypomonagreen}{rgb}{0.12,0.3, 0.17}
\definecolor{darkpastelgreen}{rgb}{0.01, 0.75, 0.24}
\definecolor{green-yellow}{rgb}{0.68, 0.98, 0.68}
\definecolor{olive}{rgb}{0.5, 0.5, 0.0}
\definecolor{kellygreen}{rgb}{0.3, 0.73, 0.09}
\definecolor{lava}{rgb}{0.81, 0.06, 0.13}
\definecolor{mordantred19}{rgb}{0.68, 0.05, 0.0}
\definecolor{pastelred}{rgb}{1.0, 0.41, 0.38}
\definecolor{persianred}{rgb}{0.8, 0.2, 0.2}
\definecolor{red-brown}{rgb}{0.65, 0.16, 0.16}
\definecolor{scarlet}{rgb}{1.0, 0.13, 0.0}
\definecolor{rossocorsa}{rgb}{0.83, 0.0, 0.0}
\definecolor{tangelo}{rgb}{0.98, 0.3, 0.0}
\definecolor{safetyorange}{rgb}{1.0, 0.4, 0.0}
\definecolor{pastelgreen}{rgb}{0.47, 0.87, 0.47}
\definecolor{palegreen}{rgb}{0.6, 0.98, 0.6}
\definecolor{internationalorange}{rgb}{1.0, 0.31, 0.0}

\begin{document}

\title{Quantum Fisher Information and multipartite entanglement in spin-1 chains}
\author{Federico Dell'Anna}
\author{Sunny Pradhan}
\affiliation{Dipartimento di Fisica e Astronomia dell’Università di Bologna, I-40127 Bologna, Italy}
\affiliation{INFN, Sezione di Bologna, I-40127 Bologna, Italy}

\author{Cristian Degli Esposti Boschi}
\affiliation{INFN, Sezione di Bologna, I-40127 Bologna, Italy}
\affiliation{CNR-IMM, Sezione di Bologna, via Gobetti 101, 40129, Bologna, Italy}

\author{Elisa Ercolessi}
\affiliation{Dipartimento di Fisica e Astronomia dell’Università di Bologna, I-40127 Bologna, Italy}
\affiliation{INFN, Sezione di Bologna, I-40127 Bologna, Italy}

\begin{abstract}

In this paper, we study the ground state Quantum Fisher Information (QFI) in one-dimensional spin-$1$ models, as witness to Multipartite Entanglement.
The models addressed are the Bilinear-Biquadratic  model, the most general isotropic $SU(2)$-invariant spin-$1$ chain, and the XXZ spin-$1$ chain, both with nearest-neighbor interactions and open boundary conditions.
We show that the scaling of the QFI of strictly non-local observables can be used for characterizing the phase diagrams and, in particular, for studying topological phases, where it scales maximally.
Analyzing its behavior at the critical phases, we are also able to recover the scaling dimensions of the order parameters, both for local and string observables.
The numerical results have been obtained by exploiting the Density Matrix Renormalization Group algorithm and Tensor Network techniques.
\end{abstract}
\maketitle

\section{Introduction}%
\label{sec:introduction}

In addition to be a crucial resource for quantum-enhanced metrology \cite{pezze2014quantum} and quantum computation \cite{cirac2012goals}, entanglement has been used to characterize quantum phases and quantum phase transitions (QPTs) in many-body models, particularly for low-dimensional systems, and has been important also to uncover exotic states of matter like topological spin liquids \cite{savary2016quantum} or to describe many-body localization \cite{abanin2019localization}.

Bipartite entanglement has been the primary focus in the literature \cite{Amico2008}, with the area law \cite{Eisert2010} serving as a benchmark for relating the amount of entanglement between two partitions of a quantum many-body system to the surface area between the blocks \cite{latorre2009short,Vidal2003}.
It has been proved \cite{Guhne2005multipartite} that the ground state of some spin chains should exhibit \emph{Multipartite Entanglement} (ME), but somehow this topic has received less attention \cite{Guhne2005multipartite}, despite the fact that many-body quantum states are far more complex than what can be captured with bipartite entanglement only.

A possible estimator of multipartite entanglement is  \emph{Quantum Fisher Information} (QFI), a quantity which is introduced in the context of the problem of phase estimation in metrology \cite{giovannetti2006quantum} and is of use in the study of the sensitivity of atomic interferometers beyond the shot-noise limit \cite{Giovannetti2004quantum}.
The QFI associated to local operators has recently been used to observe ME in models exhibiting Ginzburg-Landau-type quantum phase transitions \cite{hauke2016measuring} and in spin systems such as the Ising, XY, and Heisenberg models \cite{hauke2016measuring, liu2013quantum, lambert2023heisenberg} also at finite temperature \cite{lambert2019heisenberg}, where ME is expected to diverge at criticality. It has been pointed out, however, that the use of local operators in this method fails to detect ME at topological quantum phases and transitions.
To address this issue, QFI-based methods need to be extended  to include also non-local operators, as first outlined in \cite{Pezze2017entanglement, pezze2016witnessing}.

In this paper, we are going to study the ME in two paradigmatic spin-1 systems with nearest-neighbor interactions: the \emph{Bilinear-Biquadratic (BLBQ) model}  and the \emph{XXZ model}, two models with a rich phase diagram which exhibit a topological Haldane phase.
More specifically, we will show that QFI of non-local order parameters (such as string-order parameters \cite{kennedy1992hidden}) gives indeed information about the ME of the ground state in the different phases of the models.
Then, taking also in consideration QFI of local spin observables, we are able to classify all phases of the model as well as to calculate universal critical exponents at phase transitions.

\medskip

The paper is structured as follows.
In Sec.~\ref{sec:basic_concepts}, we briefly review ME and QFI, and their relationship.
In Sec.~\ref{sec:bilinear_biquadratic_model} we discuss the BLBQ model;
after describing its phase diagram, we analyze the scaling of the QFI with respect to some selected operators.
The same is done in the last Sec.~(\ref{sec:XXZ_model}) for the XXZ model.
A summary of the obtained results is discussed in the conclusions in Sec.~\ref{sec:conclusions}, with some possible outlooks for future research.

\section{Quantum Fisher Information and Multipartite entanglement}
\label{sec:basic_concepts}

In this section, we concisely review the concepts of ME and QFI,  elucidating their relationship \cite{Hyllus2012fisher, pezze2014quantum}.

\medskip
        A pure state of $N$ particles is $k$-\emph{producible} if it can be written as:
\begin{equation}
    \ket{\psi_{k-\text{prod}}} = \bigotimes_{l=1}^{M}\ket{\psi_l}
    \label{product}
\end{equation}
where $\ket{\psi_l}$ is a state with $N_l \leq k$ particles and $M$ is the number of parties in which it is possible to split up the state so that $\sum_{l=1}^{M} N_l=N$.

A state is $k$-\emph{entangled} if it is $k$-\emph{producible} but not $(k-1)$-\emph{producible}.
Therefore, a $k$-particle entangled state can be written as a product $\ket{\psi_k}$ which contains at least one state $\ket{\psi_l}$ of $N_l=k$ particles which does not factorize further.
So, in this notation, a state $\ket{\psi_{1-\text{ent}}}$ is fully separable while a state $\ket{\psi_{N-\text{ent}}}$ is maximally entangled.
These definitions can be extended to mixed states via convex combination.

\medskip

QFI is a fundamental quantity in the context of \emph{phase estimation} and is crucial to prove that entanglement can increase the sensitivity of an interferometer beyond the shot noise up to the Heisenberg limit. QFI for a general observable $\hat{O}$ and a mixed probe state $\rho = \sum_{i} p_i \op{\phi_i}{\phi_i}$, with $p_i > 0$ and $\sum_{i} p_i = 1$, is given by
\begin{equation}
    \FQ[\rho, \hat{O}] = 2 \sum_{i,i'}  \frac{(p_i - p_{i'})}{p_i + p_{i'}} \abs{ \mel{\phi_i}{\hat{O}}{\phi_{i^{\prime}}} }^2.
\end{equation}
In the case of a pure state $\ket{\psi}$ the QFI has a simple expression and  is directly proportional to the variance of the operator:
\begin{equation}
    \FQ[ \ket{\psi}, \hat{O}] = 4 (\Delta \hat{O})^2 \equiv 4( \ev*{\hat{O}^2} - \ev*{\hat{O}}^2). \label{eq:FQ}
\end{equation}
For separable states, $\rho_{\text{sep}}$ the $\FQ[\rho_{\text{sep}}, \hat{O}]$ is bounded from above \cite{Giovannetti2004quantum}:
\begin{equation}
   \FQ[\ket{\psi}_{\text{sep}}, \hat{O}]\leq N(\lambda_{\text{max}}-\lambda_{\text{min}})
\end{equation}
where $\lambda_{\text{max}}$ and $\lambda_{\text{min}}$ are the maximum and minimum eigenvalue of $\hat{O}$.
This is not a fundamental limit, since  it can be surpassed by using proper entangled states.
Indeed, for general probe pure states $\ket{\psi}$ of $N$ particles, we have \cite{Giovannetti2004quantum, Pezze2009entanglement}
\begin{equation}
    \FQ[\ket{\psi}, \hat{O}] \leq N^2(\lambda_{\text{max}}-\lambda_{\text{min}})^2,
\end{equation}
where the equality is saturated by only maximally entangled states. This gives the Heisenberg limit in phase estimation and quantum interferometer theory.

\medskip

There is a direct relationship between ME and QFI, as it has been show in \cite{Hyllus2012fisher}.
For any $k$-producible states $\ket{\psi}_{k-\text{prod}}$  of $N$ particles, the QFI is bounded by
\begin{equation}
    F_{\text{Q}} [\ket{\psi}_{k-\text{prod}}, \hat{O}] \leq sk^2 + r^2
    \label{eq:F_Q_criterion}
\end{equation}
where $s = \floor{N/k}$ (the integer part of $N/k$) and $r = N - sk$.
Therefore, a violation of \eqref{eq:F_Q_criterion} will indicate a $(k+1)$-particle entanglement.
The quantity $\FQ$ in \eqref{eq:F_Q_criterion} has been rescaled by a factor $(\lambda_{\text{max}} - \lambda_{\text{min}})^2$, which in the case of spin-1 operators is equal to 4.
By a straightforward calculation is possible to see that this bound is saturated by the product of $s$ GHZ states of $k$ particles and a GHZ state with the remaining $r$ particles :
\newcommand{\ketmax}{\ket{\lambda_{\text{max}}}}
\newcommand{\ketmin}{\ket{\lambda_{\text{min}}}}
\begin{multline}
    \ket{\psi} = \bigotimes^s_i
    \left(
        \frac{\ketmax^{\otimes k} + \ketmin^{\otimes k }}{\sqrt{2}}
    \right)_i \times \\
    \times \left(
        \frac{\ketmax^{\otimes r} + \ketmin^{\otimes r }}{\sqrt{2}}
    \right)
\end{multline}
If we introduce the QFI density
\begin{equation}
    \fQ[\ket{\psi}, \hat{O}] \equiv \FQ[\ket{\psi}, \hat{O}] / N,
\end{equation}
then \eqref{eq:F_Q_criterion} can immediately be read as
\begin{equation}
    \fQ[\ket{\psi}_{k-\text{prod}}, \hat{O}] \leq k,
    \label{eq:f_Q_criterion}
\end{equation}
where, for simplicity, we put the term $s=N/k$.
It has been proved that $\fQ>1$ is a sufficient condition for multipartite entanglement \cite{Pezze2009entanglement}.

In this paper, the observable we consider are constructed by using the spin-1 operators $S_i^{\alpha}$, where $\alpha = x,z$, and their \emph{non-local} counterparts $\tilde{S}^{\alpha}$.
The latter are defined as follows:
\begin{equation}
    \begin{split}
        \widetilde{S}^x_j
        = S_j^x \left(e^{i\pi \sum_{l>j} S_l^x } \right), \\
        \widetilde{S}^z_j
        = \left( e^{i\pi \sum_{l<j} S_l^z } \right) S_j^z.
    \end{split}
\end{equation}
These operators have been obtained by applying a non-local unitary transformation on the spin degrees of freedom.
For more details regarding the origin of this transformation, we refer to the discussion about the AKLT model in Appendix~\ref{sec:the_aklt_model}.

\section{Bilinear-Biquadratic model}%
\label{sec:bilinear_biquadratic_model}

\begin{figure}[t]
    \centering
    \includegraphics{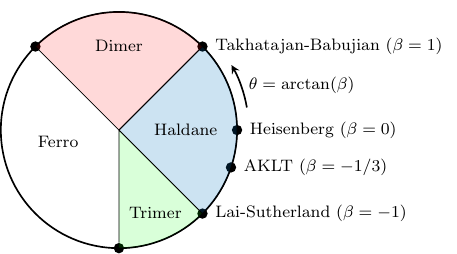}
    \caption{%
        Phase diagram of BLBQ model on a circle, parametrized by $\theta$, with some remarkable points:
        the AF Heisenberg model the AKLT point, and the critical points (Takhtajan-Babujian and Lai-Sutherland).
        In terms of $\beta$ and $J$, the right half corresponds to a positive $J$ and the left half to a negative $J$, while $\beta = \tan \theta$.
    }
    \label{fig:blbq_phase_diagram}

\end{figure}

In this section we consider the \emph{Bilinear-Biquadratic} (BLBQ) model on a chain of $N$ sites:
\begin{equation}
    H = J\sum_{i=1}^N \qty[ \bm{S}_i \cdot \bm{S}_{i+1} -\beta(\bm{S}_i \cdot \bm{S}_{i+1})^2 ],
    \label{eq:blbq_hamiltonian}
\end{equation}
where $\bm{S}_i = (S_i^x, S_i^y, S_i^z)$ is the spin-1 operator for site $i$, $J$ is the nearest-neighbor coupling and $\beta$ is a real parameter expressing the ratio between the bilinear and biquadratic terms. This is the most general $SU(2)$-invariant isotropic spin-$1$ Hamiltonian with nearest-neighbor interactions only.
Often in literature the Hamiltonian \eqref{eq:blbq_hamiltonian} is written as
\begin{equation}
    H = J^{\prime} \sum_{i=1}^N \qty[ \cos(\theta)\bm{S}_i \cdot \bm{S}_{i+1} -\sin(\theta)(\bm{S}_i \cdot \bm{S}_{i+1})^2 ],
    \label{eq:blbq_hamiltonian_alt}
\end{equation}
which can be obtained by setting $ J = J^\prime\cos(\theta) $ and $\beta = \tan(\theta)$, with the angular parameter $\theta \in [-\pi, \pi]$.
By fixing $J^{\prime} = 1$, the phase diagram can be drawn by varying the angular parameter $\theta$, as shown in Fig.~\ref{fig:blbq_phase_diagram}.

In the following we will describe the phases of the BLBQ model and some remarkable points.

\subsection{Phase Diagram}
\label{sub:BLBQ_model}

\begin{figure}[t]
    \centering
    \includegraphics{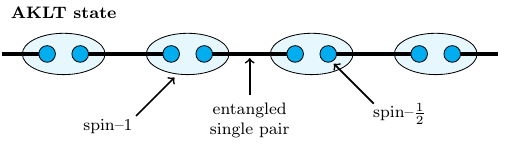}

    \vspace*{0.3cm}
    \includegraphics{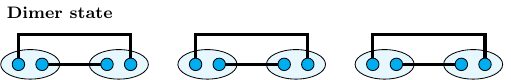}
    \caption{
        Example of states of the BLBQ model: every site (light blue oval) contains two spin-$1/2$ particles (blue dots) and each singlet state of spin-$1/2$ particles is represented with a thick black line.
        \emph{Upper panel.}
        Entangled pair structure of the AKLT's ground state in the VBS representation.
        \emph{Lower panel.}
        An example of spin-1 dimer state in the VBS representation with six sites.
    }
    \label{fig:blbq_states}
\end{figure}

\medskip
The \emph{Haldane phase} corresponds to the region  $-1 < \beta < 1$ and $J > 0$: here the system is massive, with a unique ground state and exponentially decaying correlation functions \cite{affleck1988valence}. We recognize the antiferromagnetic Heisenberg model for $\beta = 0$ \cite{Haldane1983continuum, Haldane1983nonlinear,nightingale1986gap}.
For $\beta = -1/3$ we recover the AKLT model, whose ground state is a Valence-Bond State (VBS), in which each spin-1 is thought of as made of two $1/2$-spins that couple with the spins of neighboring sites in a singlet (entangled) state.
A pictorial image of the AKLT state for a four sites chain is given in the upper panel of Fig.~\ref{fig:blbq_states}.

The ground state has an exact description as a Matrix Product State, which is very useful for performing exact calculations.
In particular, it can be shown that the local correlation functions have an exponential decay (see Appendix~\ref{sec:the_aklt_model}).

\medskip
The \emph{Dimer phase} corresponds to $\beta > 1$ and $J>0$, or $\beta < -1$ and $J<0$:
the system has a two-fold degenerate ground state and a small excitation gap \cite{barber1989spectrum}.
The degeneracy is due to the broken translation symmetry, since neighboring spins tend to be coupled in pairs. A good approximation of the ground state in the whole phase is given by the dimer state \cite{kennedy1992hidden}:
\begin{multline}
    \ket{d}_{\pm} =
    \bigotimes_{i=1}^{L/2} \frac{1}{\sqrt{3}}
    \Big(
        \ket{+}_{2i} \ket{-}_{2i \pm 1} + \\ +
        \ket{-}_{2i} \ket{+}_{2i \pm 1} -
        \ket{0}_{2i} \ket{0}_{2i \pm 1}
    \Big)
    \label{eq:dimer}
\end{multline}
which is shown in the lower panel of Fig.~\ref{fig:blbq_states}.
Haldane and Dimer  phases are separated by the so-called \emph{Takhtajan-Babujian critical point}, for $\beta = 1$ and $J > 0$.
Here the Hamiltonian is integrable by means of Bethe Ansatz technique \cite{takhtajan1982picture,babujian1982exact} and its universality class is that of a $SU(2)_k$ Wess-Zummino-Witten conformal field theory with  $k = 2$ and therefore with central charge $c = 3/2$ \cite{Francesco1997}.

\medskip
In the region $\beta < -1$ and $J > 0$ there is another antiferromagnetic phase, called the \emph{Trimer Phase}, since the ground state tends to be invariant under translations of three sites.
This is a gapless phase \cite{solyom1987competing}.
At $\beta = -1$, it is separated from the Haldane phase by a continuous phase transition.
This point corresponds to the so-called \emph{Lai-Sutherland model}, which has an enhanced symmetry to $SU(3) $, the Hamiltonian being equivalent to
\begin{equation}
    \sum_{i=1}^{N-1}  \bm{S}_i \cdot \bm{S}_{i+1} +(\bm{S}_i \cdot \bm{S}_{i+1})^2 = \frac{N}{3} +\frac{1}{2}\sum_{i=1}^{N-1} \sum_{a=1}^{8}\lambda_i^a
\end{equation}
where $\lambda^a$ are the Gell-Mann matrices, the eight generators of $SU(3)$ algebra.
It is in the universality class of the $SU(3)_k$ Wess-Zummino-Witten conformal field theory with  $k = 1$ \cite{Francesco1997, sutherland1975model}.
Here we will not consider the last phase present in Fig.~\ref{fig:blbq_phase_diagram}, namely the ferromagnetic phase, which corresponds to an ordered and separable ground state.

\begin{figure*}[t]
    \centering
    \subfloat[Haldane phase.\label{Scaling_Haldane}]{
        \includegraphics{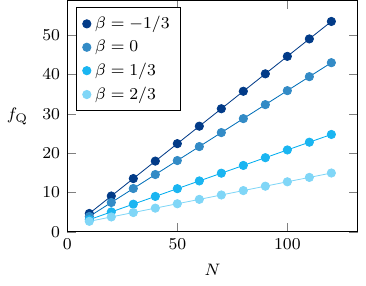}
    }
    \subfloat[Dimer and trimer phase.\label{Scaling_Dimer}]{
        \includegraphics{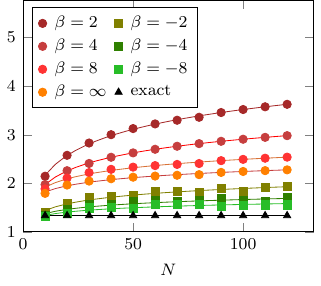}
    }
    \subfloat[QFI density at $N=30$.\label{var_beta}]{
        \includegraphics{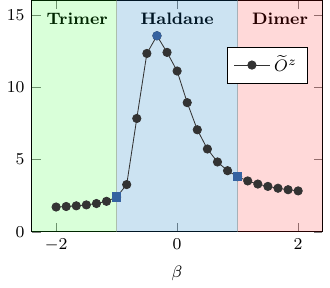}
    }
    \caption{
        Scaling behaviors of the QFI density $\fQ$ at different points of phase diagram using $\tilde{O}^z$ in \eqref{eq:string_op}:
        (a) in the Haldane phase $\fQ$ grows linearly, with the highest slope in correspondence of the AKLT point $\beta = -1/3$;
        (b) in the dimer and trimer phase $\fQ$ grows logarithmically;
        (c) values of $\fQ$ as a function of $\beta$ with system size $N=30$; the blue circle is the AKLT point where $\fQ$ is maximal, while the two blue squares correspond to the phase-transition points.
    }
\end{figure*}

The BLBQ model has a hidden symmetry (see Appendix), that  forces to introduce \emph{non-local order parameters} (NLOPs) \cite{kennedy1992hidden} to classify all phases.
NLOPs, which are also called String Order Parameters, are defined as follows:
\begin{equation}
    \tilde{C}^\alpha = \lim_{r \to \infty}\bigg\langle S^{\alpha}_1\bigg( \prod_{k=2}^{r-1} e^{i \pi S_k^{\alpha}}\bigg) S^{\alpha}_r \bigg\rangle
    \label{NLOP}
\end{equation}
where $\alpha=x,y,z$.
The NLOPs $\tilde{C}^{\alpha}$ have a non-zero expectation value only in the Haldane phase.

In the following, we will examine both the expectation value and the QFI of the non-local operator:
\begin{equation}
  \tilde{O}^z\equiv \sum_{j=1}^N  \widetilde{S}_j^z \;\; , \;\;  \widetilde{S}^z_j  \equiv  \left( e^{i\pi \sum_{l<j} S_l^z } S_j^z \right)
  \label{eq:string_op}
\end{equation}
evaluated on the ground state $|\psi\rangle$ in the different phases of the BLBQ model. With some algebra one finds:
\begin{equation}
    \ev*{\tilde{O}^z}
    = \sum_{l = 1}^{N}
        \Big\langle
            \bigg( \prod_{j = 2}^{l-1} \Omega(j) \bigg) S_l^z
        \Big\rangle
    \label{eq:secondterm}
\end{equation}
and
\begin{equation}
        \begin{split}
            \ev*{(\tilde{O}^z)^{2}} =
            \sum_{l = 1}^{N} \ev*{ (S^z_l)^2 }
            - 2 \sum_{l < m}
                \Big\langle
                    S^z_l \bigg( \prod_{j = l+1}^{m-1} \Omega(j) \bigg) S^z_m
                \Big\rangle,
        \end{split}
    \label{first_term}
\end{equation}
where we have used
\begin{equation}
    \Omega(l) =  e^{i \pi S_l^z}, \quad
    \Omega ^2 (l)  = \mathbb{I} \quad \text{and} \quad S_l^z \Omega (l)  = -S_l^z.
\end{equation}
These expressions are used to calculate the QFI
\begin{equation}
    F_Q\big[\ket{\psi},\tilde{O}^z \big] = \bigg[\bra{\psi} (\tilde{O}^z)^{2}\ket{\psi}-\bra{\psi} \tilde{O}^z\ket{\psi}^2\bigg]
    \label{Fvariance}
\end{equation}
which coincides with (\ref{eq:FQ}) but for the factor $4$ that we have neglected since we are dealing with spin-1 operator with $\lambda_{\text{max}}=-\lambda_{\text{min}}=1$.

\subsection {Numerical results}
\label{subsec:BLBQ_results}

To rewrite \eqref{first_term}, it is useful to define the following $N \times N$ matrix:
\begin{equation*}
    M =
    \left(
        \begin{smallmatrix}
            \ev*{(S_1^z)^2} & \ev*{S_1^z S_2^z} & \ev*{S_1^z \Omega(2) S_3^z} & \cdots & \ev*{S_1^z \Omega(2)\cdots \Omega(N-1) S_N^z} \\
            0               & \ev*{(S_2^z)^2}   & \ev*{S_2^z S_3^z}           & \cdots & \ev*{S_2^z \Omega(3)\cdots \Omega(N-1) S_N^z} \\
            0               & 0                 & \ev*{(S_3^z)^2}             & \cdots & \cdots                                        \\
            \cdots          & \cdots            & \cdots                      & \cdots & \cdots                                        \\
            0               & 0                 & 0                           & \cdots & \ev*{S_{N-1}^z S_N^z}                         \\
            0               & 0                 & 0                           & \cdots & \ev*{(S_N^z)^2}
        \end{smallmatrix}
    \right),
\end{equation*}
where each matrix element $M_{ij}$ is given by
\begin{equation}
    M_{ij} =
    \begin{cases}
        \ev*{S_i^z \, \Omega(i+1) \cdots \Omega(j-1) \, S_j^z} & \text{if $i \leq j$} \\
        0 & \text{otherwise.}
    \end{cases}
\end{equation}
Similarly, for the term \eqref{eq:secondterm} we can define the $N$-dimensional vector:
\begin{equation}
    V =  \left( \ev*{ S_1^z}, \ev*{ \Omega(1)S_2^z}, \dots, \ev*{ \Omega(1) \cdots \Omega(N-1))S_N^z} \right),
\end{equation}
such that $\ev*{\tilde{O}^z}$ turns out to be the sum of all its elements.

In this way, the QFI can be written as
\begin{equation}
    \label{eq:Fisher_M_V}
    \FQ \qty[\ket{\psi},\tilde{O}^z] =
    \sum^{N}_{i=1}M_{ii}  -2 \sum^{N-1}_{i=1} \sum^{N}_{j>i} M_{ij} - \qty( \sum_{i=1}^{N}  V_i )^2,
\end{equation}
Simulations to compute the elements of $M$ and $V$ can be easily implemented numerically.
The states can be represented with Matrix Product States (MPSs) and the ground states can be obtained with the DMRG algorithm.
The numerical simulations have been done using the ITensor library \cite{Itensor,itensor-r0.3} and the DMRG computations have been performed with bond dimensions up to $\chi = 300$ and truncation error cutoff set to $10^{-12}$, for a higher precision.

In order to investigate the scaling of the QFI density $\fQ= \FQ/N$, we have looked for a function of the form $q+bN^\delta$ (for the Haldane and critical points) or $q+b \ln N$ (for the dimer and trimer phases), for system sizes up to $N= 120$.
However, when the data showed a particularly flat trend, we have fitted $\fQ$ against a constant function, in order to minimize the standard error on the parameters.

The results of the numerical calculations are summarized in Table~\ref{table_scaling_haldane} for the Haldane  phase and in Table~\ref{table_scaling_dimer} for the Dimer and Trimer phases.
The fit and their errors are computed using standard methods, like the one provided by Mathematica \cite{Mathematica}.

To analyze these results, let us start from the AKLT point, where the ground state is known exactly.
To calculate the QFI analytically, we can exploit Lemma 2.6 of \cite{affleck1988valence}, extended to a string observable.
Let $O$ be an observable and $N$ the system's size; then for any $l \leq N$ such that the support of $O$ is contained in $l$, we have
\begin{equation}
    \lim_{N \to \infty}
    \frac{
        \ev*{O}{\Omega^N_{\alpha \beta}}
    }{
        \braket*{\Omega^N_{\alpha \beta}}{\Omega^N_{\alpha \beta}}
    }
    =
    \frac{
        \sum_{\alpha,\beta}\ev*{O}{\Omega^l_{\alpha \beta}}
    }{
        \sum_{\alpha,\beta}\braket*{\Omega^l_{\alpha \beta}}{\Omega^l_{\alpha \beta}}
    }
\end{equation}
where $\ket*{\Omega^N_{\alpha \beta}}$ is one of the four ground states of the AKLT model (see Appendix~\ref{sec:the_aklt_model}).
This gives us an operational way to analytically calculate the terms of the QFI on the infinite volume ground state from \eqref{eq:Fisher_M_V} for a finite chain.
It turns out that each diagonal term is equal to $2/3$ while each of the $N(N-1)/2$ off diagonal terms quickly approach to $-4/9$ (i.e. the value of NLOP \eqref{NLOP} defined in the asymptotic limit) when $N$ becomes larger.
As the last addend in \eqref{eq:Fisher_M_V} is negligible, the QFI density for a system of $N$ sites scales linearly as:
\begin{equation}
    \fQ(\ket{\psi_{\text{AKLT}}},\tilde{O}^z)\simeq \frac{2}{9} + \frac{4}{9}N
\end{equation}
as confirmed by numerical results in Table \ref{table_scaling_haldane}.
The same argument holds for the Heisenberg point, where the asymptotic value of its NLOP is known to be $\simeq 0.36$ \cite{degli2005low}.
Furthermore, we observe that the QFI keeps a linear scaling in the whole Haldane phase, as shown in Fig.~\ref{Scaling_Haldane}.
One can notice that the slope of the curves progressively decreases as we move away from the AKLT point.

When moving outside the Haldane phase, the linear scaling in the dimer and trimer phase becomes sublinear, as it can be seen in Fig.~\ref{Scaling_Dimer}.
In the dimer phase, the numerical results can be compared with the analytical calculations performed on the dimer state \eqref{eq:dimer} which can be considered a good approximation,
as mentioned in Sec.~\ref{sec:bilinear_biquadratic_model}.
The resulting QFI density $\fQ(\ket{d},\tilde{O}^z)$ yields $4/3$, corresponding to a $2$-partite entanglement structure, which is expected from the state \eqref{eq:dimer} being a two-sites product state.
Then, assuming that $\tilde{O}^z$ is a good choice for the whole dimer phase, we can appreciate how good this approximation is in the different points of this phase, by comparing the various scaling with the exact value $4/3$.
As we show in Table \ref{table_scaling_dimer} and Fig.~\ref{Scaling_Dimer}, we get that a good function that fits the data is of the form $q+b\log{N}$, with $b$ that progressively decreases when $\beta$ goes to infinity.

We want to stress the crucial difference between the Haldane phase and the dimer and trimer ones.
From the point of view of QFI criterion, the multipartite entanglement structure, in other words the $k$ in \eqref{eq:f_Q_criterion}, grows linearly with the system size in the Haldane phase while in the other two phases the $k$ grows sublinearly.
This may suggest that the ground state in the Haldane phase may not be factorizable in blocks of finite length in the thermodynamic limit, and this can be shown using only \emph{non-local operators}.
However, we cannot have direct information on the exact value of $k$ using only $\tilde{O}^z$,
because we cannot be sure that this is the operator saturating the ground state QFI.

\medskip

\begin{figure}[t]
    {\centering\footnotesize \textbf{Takhtajan-Babujian} ($\beta=1$)} \\[4pt]
    \includegraphics{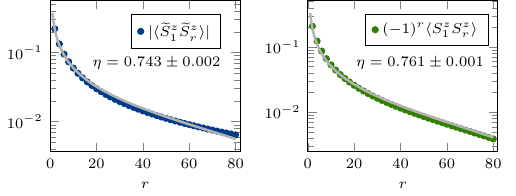}
    {\centering\footnotesize \textbf{Lai-Sutherland} ($\beta=-1$)} \\[4pt]
    \includegraphics{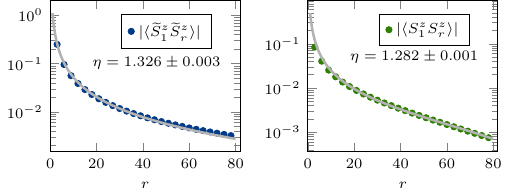}
    \caption{
        Power-law decay of correlation functions by using both string (in blue) and local (in green) operators in the Takhtajan-Babujian (upper panel) and Lai-Sutherland (lower panel) models.
    The dots are the computed value, while the gray line is obtained fit.
    }
    \label{fig:correlators}
\end{figure}

Let us now analyze the scaling behavior at the transition points $\beta = \pm 1$.
The spin-spin correlations are asymptotically given by the fundamental WZW primary fields, leading to the prediction that, in an infinite system, the dominant antiferromagnetic correlations decay as a power law:
\begin{equation}
    \langle S^{\alpha}_0 S^{\alpha}_r \rangle \sim \frac{(-1)^r}{\abs{r}^{\eta}}
    \label{eq:corr}
\end{equation}
where $\eta=2\Delta$ and the scaling dimension $\Delta=h+\bar{h}$ can be obtained from the primary field scaling dimension for a general $SU(n)$ level $k$ WZW model \cite{knizhnik1984current}:
\begin{equation}
    h=\bar{h}=\frac{n^2-1}{2n(n+k)}
    \label{eq:conformal_h}
\end{equation}
As we said in the previous sections, $\beta=\pm 1$ are described by $SU(2)_2$ and $SU(3)_1$ conformal theories which means that their values of $\eta$ are equal to $3/4$ and $4/3$ respectively.
We recover this power-law scaling of correlators both for string and local operators, as we show in Fig.~\ref{fig:correlators}.

For $\beta=1$, the numerical data display small oscillations between $N$ even and odd, due to the double degeneracy that emerges in the dimer phase.
To increase the accuracy of the fitting, we have decided to consider only the odd-numbered sites, this however does not modify the value of the exponents in the thermodynamic limit since these oscillations tend to zero as $N$ increases.

As shown in \cite{hauke2016measuring}, the QFI density of one-dimensional models at the critical point is supposed to scale as $f(O^{\alpha}) \sim N^{\delta_{Q}}$ (up to a non-universal pre-factor and sub-leading corrections) with $\delta_{Q}=1-2\Delta^{\alpha}$, where $\Delta^{\alpha}$ the scaling dimension of the operator $O^{\alpha}$. We can recover this result from our approach and numerical data as well. Indeed, considering that the first sum in \eqref{eq:Fisher_M_V} goes as $\sim N$ (so it brings just a constant contribution in $f$) and neglecting $V$ (because we are at the critical point), the only relevant contribution is given by the sum of the off-diagonal terms in the $M$ matrix.
Exploiting the \eqref{eq:corr} in the continuum limit, we get:
\begin{equation}
    \sum_{r'=1}^{N-1}\sum_{r>r'}^{N}\langle S^{\alpha}_{r'} S^{\alpha}_r \rangle \longrightarrow \int_{1}^{N}dr' \int_{r'}^{N}   \frac{dr}{r^{\eta}} \sim N^{2-\eta}
\end{equation}
so that:
\begin{equation}
    \fQ(O^\alpha)\sim N^{1-2\Delta^{\alpha}}
\end{equation}
The same holds for string operators up to a non-universal pre-factor and sub-leading corrections.
It is evident now why we get the expected numerical value $\delta\simeq \delta_{Q}=1-2\Delta= 1/4$ for the string magnetization, as reported in Table~\ref{table_scaling_haldane}.
A similar reasoning can be put forward for the calculation of  $\fQ(O^z_{\text{st}})$ of the local staggered magnetization operator along $z$-axis, defined as
\begin{equation}
    O^z_{\text{st}} = \sum_{j=1}^{N} (-1)^{j} S_j^z.
    \label{eq:stag_magn}
\end{equation}
Our numerical results for the calculation of the QFI density for $O^z_{\text{st}}$ yield:
$q = \num{-3.770+-0.002}$, $b = \num{3.201+-0.001}$ and $\delta = \num{0.244+-0.001}$.
Thus, we are able to read the critical exponent of the operator from its QFI.

At the Lai-Sutherland point $\beta=-1$, the numerical data display small oscillations with a periodicity of three sites, due to the trimer configuration that merges for $\beta <-1$.
Unfortunately, from the data we observe what is mostly probable a flat trend, but  we are not able to distinguish a linear fit from a one that decreases exponentially or, like it should be in this case, as a power law with a negative exponent.
We believe that the pre-factors and sub-leading terms, that depend on $N$, might contribute to mask the predicted behavior at criticality.

\begin{table}
    \setlength{\tabcolsep}{5pt}
    \renewcommand{\arraystretch}{1.15}
    \sisetup{separate-uncertainty, table-format=1.4(1), table-align-uncertainty, tight-spacing}
    \begin{tabular}{r S[table-format=-1.3(1)] S S }
        \toprule \addlinespace[5pt]
        \multicolumn{4}{c}{BLBQ model, Haldane phase} \\
        \multicolumn{4}{c}{$\fQ(\ket{\psi_{\beta}}, \tilde{O}^z) = q + bN^{\delta}$} \\[4pt]
        \midrule
        $\beta$ & $q$ & $b$ & $\delta$ \\
        \midrule
        $-1/3$ & 0.225+-0.003  & 0.4441+-0.0001 & 1.0002+-0.0001 \\
        $0   $ & 0.35+-0.05    & 0.355+-0.002   & 1.002+-0.003   \\
        $1/3 $ & 1.122+-0.009  & 0.197+-0.004   & 0.9999+-0.0005 \\
        $2/3 $ & 1.55+-0.04    & 0.111+-0.02    & 0.997+-0.001   \\
        $1   $ & -3.632+-0.004 & 3.132+-0.002   & 0.252+-0.001   \\
        \bottomrule
    \end{tabular}
    \caption{Numerical values of the fitting parameters at different points in the Haldane phase of the BLBQ model, where the QFI density has been fitted against a power law.}
    \label{table_scaling_haldane}

    \vspace*{0.5cm}
    \setlength{\tabcolsep}{3.5pt}
    \renewcommand{\arraystretch}{1.15}
    \sisetup{separate-uncertainty, table-format=1.2(1), table-align-uncertainty, tight-spacing}
    \begin{tabular}{cSS[table-format=1.3(1)] | cSS}
        \toprule \addlinespace[5pt]
        \multicolumn{6}{c}{BLBQ model} \\
        \multicolumn{6}{c}{$f(\ket{\psi_{\beta}},\tilde{O}^z)=q+b\ln N$}\\[4pt]
        \midrule
        \multicolumn{3}{c|}{Dimer phase} &
        \multicolumn{3}{c}{Trimer phase} \\
        $\beta$ & $q$ & $b$ & $\beta$ & $q$ & $b$ \\[2pt]
        \hline
        $2$      & 0.81+-0.04 & 0.58+-0.01    & $-2$ & 0.98+-0.06 & 0.19+-0.01 \\
        $4$      & 1.03+-0.01 & 0.405+-0.003  & $-4$ & 1.08+-0.06 & 0.12+-0.01 \\
        $8$      & 1.34+-0.05 & 0.24+-0.01    & $-8$ & 1.11+-0.05 & 0.09+-0.01 \\
        $\infty$ & 1.39+-0.04 & 0.18+-0.01   \\
        \bottomrule
    \end{tabular}
    \caption{Numerical values of the fitting parameters at different points in the dimer and trimer phases of the BLBQ model, where the QFI density has been fitted against a logarithm function.}
    \label{table_scaling_dimer}
\end{table}

\section{XXZ spin-1 model}%
\label{sec:XXZ_model}

\subsection{Phase diagram}%

The XXZ spin-1 chain  is a well-studied quantum system that exhibits an interesting phase diagram as a function of the anisotropy parameter $J_{z}$.
It has the following Hamiltonian:
\begin{equation}
    H = \sum_{i=1}^{N-1} J_{xy}(S^x_i S^x_{i+1}+ S^y_i S^y_{i+1}) + J_z(S^z_i S^z_{i+1})
    \label{xxz_hamiltonian}
\end{equation}
where we take $J_{xy}=1$ and let $J_z$ vary.
It can also be considered as a particular case of the so-called $\lambda-D $ model \cite{kennedy1992hidden}, that includes also an isotropy term of the form $\sum_{i=1}^N D(S^{z}_{i})^2$.

The quantum phase diagram of this Hamiltonian has been extensively studied \cite{kitazawa1996phase}.
It includes the \emph{Haldane phase} for $0< J_z \sim 1$.
A second-order phase transition occurs from the Haldane phase to an antiferromagnetic (AFM) phase that belongs to the same universality class of the 2D Ising model with central charge $c=1/2$.
Various numerical techniques, including Monte-Carlo \cite{nomura1989spin} and DMRG \cite{heng2012non, liu2014entanglement}, have determined the critical value:  $J_z^{(\text{IS})}=1.186$.
A \emph{Berezinskii-Kosterlitz-Thouless} (BKT) transition occurs at $J_z^{(\text{BKT})} =0$ between the Haldane phase and a gapless disordered XY phase ($-1<J_z<0$).
The value of $J_z^{(\text{BKT})}$ is theoretically predicted to be exactly zero, using bosonization techniques \cite{schulz1986phase}.
Numerically, this has been verified via finite-size scaling \cite{ueda2008finite,botet1983ground} and DMRG \cite{heng2012non}.
The entire XY phase (including the BKT transition point) is a critical phase, which has conformal symmetry with central charge $c=1$.
Finally, at $J_z = -1$, a first-order phase transition from the XY phase to a ferromagnetic (FM) phase takes place \cite{kitazawa1996phase, liu2014entanglement, chen2003ground}.
We will not examine in detail such ferromagnetic phase in the following.

\begin{figure}
    \includegraphics{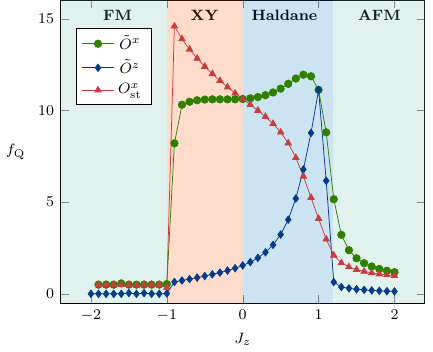}
    \caption{
        Trend of QFI densities in the phase diagram of the XXZ model of size $N=30$, with the string magnetizations $\tilde{O}^z$, $\tilde{O}^x$, and the local magnetization $O^x_{\text{st}}$.
        From left to right: ferromagnetic (FM) phase, XY gapless phase, Haldane phase and antiferromagnetic (AFM) phase.
        The critical points are located at $J_z = -1$, $J_z^{(\text{BKT})} = 0$ and $J_z^{(\text{IS})} = 1.186$.
    }
    \label{opers_in_phasesXXZ}
\end{figure}

\subsection{Numerical results}

\begin{figure*}[t]
    \subfloat[Staggered magnetization $O^x_{\text{st}}$\label{sx_stagg_N}]{
        \includegraphics{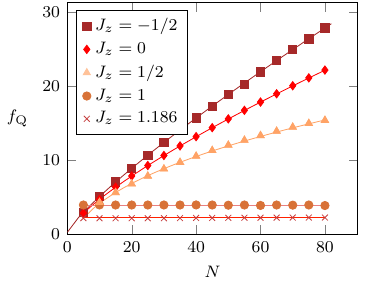}
    }
    \subfloat[String operator $\tilde{O}^x$\label{Sx_N}]{
        \includegraphics{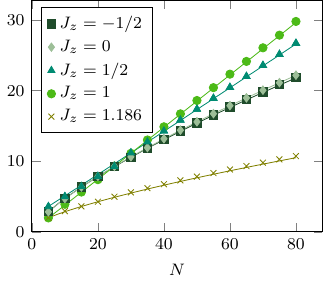}
    }
    \subfloat[String operator $\tilde{O}^z$\label{Sz_N}]{
        \includegraphics{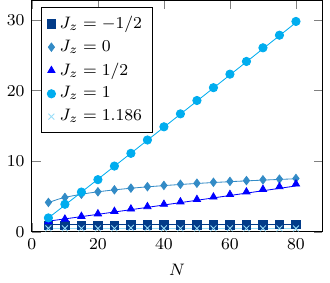}
    }
    \caption{
        Scaling behaviors of the QFI density of different operators at some points of interest:
        (a) staggered magnetization $O^x_{\text{st}}$ (shades of red);
        (b) $x$-string operator $\tilde{O}^x$ (shades of green);
        (c) $z$-string operator $\tilde{O}^z$ (shades of blue).
        Notice the abrupt change in behavior for the string operators $\tilde{O}^x$ and $\tilde{O}^z$ from $J=1$ to $J=1.186$, which can also be seen in Fig.~\ref{opers_in_phasesXXZ}.
    }
    \label{fig:qfi_density_xxz}
\end{figure*}

Given the symmetries of the Hamiltonian, we consider the scaling behavior of the QFI density of local and string operators along the $x$ and $z$ axes, including the staggered ones.
The ones that show an extensive scaling, at least in some phases of the model, are the following:
\begin{equation}
    \tilde{O}^z=\sum_{i=1}^N \tilde{S}_i^z ,\quad \tilde{O}^x=\sum_{i=1}^N \tilde{S}_i^x ,\quad O^{x}_{\text{st}}=\sum_{i=1}^N (-1)^iS_i^x,
    \label{relevant_op}
\end{equation}
where, as usual, the operators with the tilde symbol are string operators.
Similarly to the previous section, the numerically computed QFI density $\fQ$ is fitted against the function $\fQ=q+bN^{\delta}$, or with a constant if the data presents an extremely flat trend.

In Fig.~\ref{opers_in_phasesXXZ} we plot the shapes of the QFI densities of the operators \eqref{relevant_op} in the different phases of the model for a chain with $N=30$ sites.
The results of the fitting of the scaling with $N$ are given in Tables \ref{table_scaling_Oxs}, \ref{table_scaling_OX} and \ref{table_scaling_OZ} and some details of the scaling are reported in Fig.~\ref{opers_in_phasesXXZ}.
Let's analyze each operator below.

\medskip

The operator $O^{x}_{\text{st}}$ takes its maximal value close to the FM-XY transition point and then decreases progressively moving toward the Haldane phase.
In particular, analyzing its scaling with $N$ (see Fig.~\ref{sx_stagg_N} and Table~\ref{table_scaling_Oxs}), $\fQ$ reveals a power-law behavior in the XY phase with the coefficient $\delta =0.8376 \pm 0.0001$ at $J_z = -1/2$ which gradually reduces (e.g. $\delta = 0.7574 \pm 0.0002$ at $J_z=0$) until it vanishes for $J_z \gtrsim 1$.

Regarding the string operators (see Tables \ref{table_scaling_OX} and \ref{table_scaling_OZ}), it is possible to observe that $\fQ(\tilde{O}^x)$ has a power-law scaling in the whole XY phase (including $J_z=0$) where the  $\fQ(\tilde{O}^z)$ appears to be almost flat, ($\delta=0.138 \pm 0.003$).
In the Haldane phase, the QFI for both these operators shows a linear scaling ($\delta\simeq 1$) with a slope that increases with $J_z$, reaching the maximal values at $J_z=0.8$ and $J_z=1$ respectively.
For $J_z=1$ we recover the Heisenberg model where both have the same scaling coefficients as expected in an isotropic point.

\medskip
The data on QFI can be used to extract information about the critical exponents of relevant operators at phase transition points and about correlation functions in general.
At the critical point $J_z^{(\text{IS})}$, we predict that the scaling dimension of the order parameter is $\Delta=1/8$, in accordance with the universality of the 2D Ising model,  since $\delta=1-2\Delta\simeq 3/4$.
This holds true for the string order operator $\tilde{O}^x$, see Table~\ref{table_scaling_OX}, and the local staggered magnetization $O^z_{\text{st}}$.
The latter is defined similarly to $O^x_{\text{st}}$ in \eqref{relevant_op}, for which we obtained $\delta = 0.76 \pm 0.01$.

More generally, we can consider the asymptotic behavior of local staggered and string correlation functions
\begin{equation}
    \begin{split}
        C^{\alpha}_{\text{st}}(r) & = (-1)^r \ev*{S_1^{\alpha} S_r^{\alpha}}, \\
        \tilde{C}^\alpha(r) & = \left\langle S^{\alpha}_1\bigg( \prod_{k=2}^{r-1} e^{i \pi S_k^{\alpha}}\bigg) S^{\alpha}_r \right\rangle
    \end{split}
\end{equation}
which are known to have the following behavior for large $r$ in the (massive) Haldane phase \cite{boschi2009effective}:
\begin{equation}
    C^{\alpha}=a_{0} \frac{ e^{-\frac{r}{a_1}}}{\sqrt{r}}, \qquad \tilde{C}^\alpha=a_2+ a_{0} \frac{ e^{-\frac{r}{a_1}}}{r^2}
    \label{eq:in Haldane}
\end{equation}
where $a_0$, $a_1$ and $a_2$ are fitting parameters and $\alpha=x,z$ as usual, while at the transition point, they scale algebraically:
\begin{equation}
    C^{z}=\tilde{C}^{x}= \frac{a_{0}}{r^{1/4}}, \quad
    C^{x}=a_{0} \frac{ e^{-\frac{r}{a_1}}}{r^{1/4}}, \quad \tilde{C}^z=a_2+  \frac{a_{0}}{r^2}.
    \label{eq:at the trans}
\end{equation}

The data reported in the Tables \ref{table_scaling_Oxs}, \ref{table_scaling_OX} and \ref{table_scaling_OZ} and Fig.~\ref{fig:qfi_density_xxz} of the fitting parameters of $\fQ$ are in agreement with these theoretical predictions. In order to understand the results, two comments are necessary.

The first one is that in the Haldane phase and at the critical points the only relevant contribution to the QFI density is due to \eqref{first_term}, i.e.~the $M$ matrix made by the spin-spin correlators.
The second one is that, as we said previously for the BLBQ, from our data it is not possible to distinguish the flat scaling of $\fQ$ from an exponential or power-law decay with $\delta<0$.
Then, considering the correlations \eqref{eq:in Haldane} and \eqref{eq:at the trans}, we can understand that for string operators in the Haldane phase, the elements $M_{ij}$ are going to approach $a_2$.
This leads to a $\fQ$ that scales linearly, with the slope $b \simeq a_2$.
From our computations  we get $\delta$ equal to $0.757\pm 0.001$ and $0.727 \pm 0.001$ for $O_{\text{st}}^z$ and $\tilde{O}^x$, respectively, which is comparable to $1 - \eta$ as expected.

Finally, when $ -1 < J_z < 0$, the system is in the XY phase.
In this extended area of critically, also called ``critical fan'', the Hamiltonian can be replaced by the Hamiltonian of a Gaussian model \cite{kohmoto1981hamiltonian}, which admits two primary operators with conformal dimensions:
\begin{equation}
    \Delta_1=\frac{1}{8}, \qquad \Delta_2=\frac{1}{4} \chi(J_z),
\end{equation}
where $\chi$ is a function of the coupling $J_z$ such that $ \chi(0)=1/2$ and $\chi(-1)=0$. The explicit  form of the function $\chi$ depends on the details about how the lattice model can be mapped to the Gaussian model at criticality.
This means that there exists one operator for which the critical index $\delta$  of QFI densities will be constantly $3/4$ and one with varying between $3/4$ and $1$, respectively. We identify such operators with $\tilde{O}^x$ and $O^x_{\text{st}}$, respectively, as it suggested by the data of
Tables \ref{table_scaling_Oxs} and \ref{table_scaling_OX}: at $J_z=0$ the values of their fitting parameters are extremely close to each other and close to $0.75$; moving toward $J_z=-1/2$, $\fQ(\tilde{O}^x)$  remains fixed to a similar value ($\delta =0.745 \pm 0.002$) while $\fQ(O^x_{\text{st}})$ has $\delta=0.8376 \pm 0.0001$ and the latter continues to increase as suggested by Fig.~\ref{opers_in_phasesXXZ}.

\begin{table}
    \setlength{\tabcolsep}{5pt}
    \renewcommand{\arraystretch}{1.15}
    \sisetup{separate-uncertainty, table-format=-1.4(1), table-align-uncertainty, tight-spacing}
    \begin{tabular}{rS[table-format=-1.3(1)]SS}
        \toprule \addlinespace[5pt]
        \multicolumn{4}{c}{XXZ model, staggered magnetization} \\
        \multicolumn{4}{c}{$f\big(\ket{\psi_{(J_z)}},O^x_{\text{st}}\big) = q+bN^{\delta}$} \\[4pt]
        \midrule
        $J_z$ & $q$ & $b$ & $\delta$ \\
        \midrule
        $-1/2 $ & 0.231+-0.003 & 0.7041+-0.0002 & 0.8376+-0.0001 \\
        $0    $ & 0.138+-0.003 & 0.797+-0.001   & 0.7574+-0.0002 \\
        $1/2  $ & -3.5+-0.5    & 2.8+-0.004     & 0.43+-0.01     \\
        $1    $ & 3.77+-0.06   &                &                \\
        $1.186$ & 2.199+-0.009 &                &                \\
        \bottomrule
    \end{tabular}

    \caption{Numerical values of the fitting parameters of the QFI density for the staggered magnetization $O^x_{\text{st}}$ at different point of the XXZ model. }
    \label{table_scaling_Oxs}

    \vspace*{0.7cm}
    \setlength{\tabcolsep}{5pt}
    \renewcommand{\arraystretch}{1.15}
    \sisetup{separate-uncertainty, table-format=1.4(1), table-align-uncertainty, tight-spacing}
    \begin{tabular}{rS[table-format=1.3(1)]SS}
        \toprule \addlinespace[5pt]
        \multicolumn{4}{c}{XXZ model, $x$-string operator} \\
        \multicolumn{4}{c}{\textbf{$f(\ket{\psi_{(J_z)}},\tilde{O}^x)=q+bN^{\delta}$}}\\[4pt]
        \midrule
        $J_z$ & $q$ & $b$ & $\delta$ \\
        \midrule
        $-1/2 $ & 0.11+-0.04   & 0.829+-0.009   & 0.745+-0.002   \\
        $0    $ & 0.131+-0.001 & 0.7992+-0.0002 & 0.7570+-0.0001 \\
        $1/2  $ & 7.2+-0.3     & 1.26+-0.03     & 0.996+-0.002   \\
        $1    $ & 0.16+-0.05   & 0.356+-0.004   & 1.008+-0.002   \\
        $1.186$ & 2.74+-0.05   & 1.65+-0.01     & 0.727+-0.001   \\
        \bottomrule
    \end{tabular}
    \caption{Numerical values of the fitting parameters of the QFI density for the $x$-string operator $\tilde{O}^x$ at different points of the XXZ model.}
    \label{table_scaling_OX}

    \vspace*{0.7cm}
    \setlength{\tabcolsep}{5pt}
    \renewcommand{\arraystretch}{1.15}
    \sisetup{separate-uncertainty, table-format=-1.3(1), table-align-uncertainty, tight-spacing}
    \begin{tabular}{rSSS}
        \toprule \addlinespace[5pt]
        \multicolumn{4}{c}{XXZ model, $z$-string operator} \\
        \multicolumn{4}{c}{$f(\ket{\psi_{(J_z)}},\tilde{O}^z)=q+bN^{\delta}$} \\[4pt]
        \midrule
        $J_z$ & $q$ & $b$ & $\delta$ \\
        \midrule
        $-1/2 $ & 0.989+-0.003 &              &              \\
        $0    $ & -3.1+-0.1    & 5.8+-0.1     & 0.138+-0.003 \\
        $1/2  $ & 1.21+-0.04   & 0.058+-0.004 & 1.03+-0.01   \\
        $1    $ & 0.16+-0.05   & 0.356+-0.004 & 1.008+-0.002 \\
        $1.186$ & 0.489+-0.007 &              &              \\
        \bottomrule
    \end{tabular}

    \caption{Numerical values of the fitting parameters of the QFI density for the $z$-string operator $\tilde{O}^z$ at different points of the XXZ model.}
    \label{table_scaling_OZ}
\end{table}

\section{Conclusions and outlooks}%
\label{sec:conclusions}

In this paper, we have shown how QFI is able to detect multipartite entanglement (ME) in spin-1 chains with short range interactions.
A key aspect in these calculations is the use of string operators whereas the QFI relative to local operators fails to detect ME, especially in the topological phases of these models, i.e.~the Haldane phase.
For the BLBQ model, given the symmetries of the Hamiltonian, we chose the string magnetization along $z$ and obtained an extensive behavior in the topological phase, signaling the divergence of ME with the system size.
The same applies to the Haldane phase of XXZ model as well.

In the dimer and trimer phases we found a sublinear behavior; in particular for the dimer phase, we also propose to use QFI density to estimate how well the 2-sites product state is approximating the various ground states in this phase.
Furthermore, we recover the expected power-law scaling of the QFI density for these 1D models in the critical phases.
In fact, by knowing the critical exponent $\eta$ of the correlators or the scaling dimension $\Delta$ of the operator with which the QFI is calculated, it is possible to predict how $\fQ$ will scale at these critical points: $\delta=1-2\Delta$.

From numerical simulation we obtained $\delta \simeq0.25$ in the Takhtajan-Babujian point of BLBQ model and  $\delta \simeq0.75$ in the AFM-Haldane transition point of XXZ model as expected.
Throughout the ``critical fan'' (XY phase) of the XXZ model, we observe a power-law behavior of $\fQ$ with two different trends of $\delta$: one fixed at the constant value of $3/4$ (string operator $\tilde{O}^x$), the other varying between $3/4$ and $1$ (staggered magnetization $O^x_{\text{st}}$) in analogy to what was done in \cite{kohmoto1981hamiltonian}.

We remark that QFI is useful for characterizing the different phases of a model, through its entanglement content.
On the other hand, it is not the most appropriate tool for localizing the transition points, because it would require a tedious analysis of how the scaling of the QFI changes close to a critical point, having to include constant terms that often complicate the fitting procedures.

In the light of these promising results, it would be interesting to investigate whether it is feasible to use it for systems with more complicated degrees of freedom, such as models with higher symmetry groups
\cite{aguado2009heisenberg} or with long range interactions \cite{Gong2016_2}.

\begin{acknowledgments}
    The authors would like to thank D.~Vodola and S.~Tibaldi for the helpful discussions.
    The work is partially supported by INFN through the project QUANTUM.
    E.E.~is also supported by the QuantERA 2020 Project QuantHEP.
\end{acknowledgments}

\appendix

\section{The AKLT model}
\label{sec:the_aklt_model}

The AKLT model is the projection point at $\beta = - 1/3$, where the Hamiltonian can be expressed as a sum over the projection operators $P_j(i,i+1)$.
Each projector acts on a pair of interacting spins for a given value of the total spin $j = 0,1, 2$.
Thus, it can be written as:
\begin{equation}
    H_{\text{AKLT}} = -\frac{2}{3}NJ +2J \sum_{i=1}^{N}P_2(i,i+1)
\end{equation}
where
\begin{equation}
    P_2(i,i+1) =
    \frac{1}{3} +
    \frac{1}{2}\qty( \bm{S}_i \cdot \bm{S}_{i+1} +\frac{1}{3}(\bm{S}_i \cdot \bm{S}_{i+1})^2 ).
\end{equation}

As shown in \cite{affleck1988valence}, the system can be thought of as made up of two spin-$1/2$ variables for each site.
By introducing the \emph{valence bond basis}, it is possible to build the ground state, called a \emph{valence bond solid} (VBS), so that in the chain there is always a bond between two neighboring spins (see upper panel of Fig.~\ref{fig:blbq_states}).

The VBS state $\ket{\text{VBS}}$ satisfies
\begin{equation}
    P_2(i,i+1) \ket{\text{VBS}} = 0 \qquad \forall i.
\end{equation}
In the spin-$1/2$'s computational basis $\psi_1 = \ket{0}$, $\psi_2 = \ket{1}$, we can construct an orthogonal basis for the $s=1$ state space, by taking the symmetrized tensor products:
\begin{equation}
    \psi_{\alpha \beta} = \frac{1}{\sqrt{2}} \qty(\psi_{\alpha} \otimes \psi_{\beta} + \psi_{\beta} \otimes \psi_{\alpha})
\end{equation}
Then, in order to contract a pair of spin-$1/2$'s to form a singlet, we use the Levi-Civita tensor of rank two:
\begin{equation}
    \Omega_{\alpha \beta} =  \epsilon^{\gamma \delta} \psi_{\alpha \gamma} \otimes \psi_{\delta \beta},
\end{equation}
where the indices $\alpha$ and $\beta$ refer to the outer spin-$1/2$'s.
It is now easy to generalize the construction for a chain of length $N$:
\begin{equation}
    \Omega_{\alpha \beta} =
    \epsilon^{\beta_1 \alpha_2} \cdots
    \epsilon^{\beta_{N-1} \alpha_N} \psi_{\alpha \beta_1} \otimes \psi_{\alpha_2 \beta_2} \otimes \cdots \otimes \psi_{\alpha_N \beta}.
\end{equation}

The AKLT model has exponentially decaying correlations, and this applies to the whole Haldane phase.
In fact, this can be shown by computing the two-point correlation function in the limit $N \to \infty$, which yields:
\begin{equation}
    \lim_{N \to \infty}\bra{\Omega}S_1^{a}S_r^{b} \ket{\Omega}
    = \delta^{ab}(-1)^r \; \frac{4}{3} \; 3^{-r}.
    \label{eq:corr_func_AKLT}
\end{equation}
showing, as anticipated, an exponentially decaying correlation function with correlation length $\xi = \ln(3)^{-1}$.
Therefore, one may conclude that there is no order in this phase but, as we will see, a different kind of hidden order is actually there.
We are going to show this fact on the valence bond state.

As it can be easily understood from Fig.~\ref{fig:blbq_states}, in a finite chain the ground state of AKLT model is four-fold degenerate due to the effective free spin-$1/2$'s at the boundaries.
Let us write the ground state of AKLT as $\Phi_{\sigma}$, where $\sigma$ is a string of $+$'s, $-$'s and $0$'s so that $\Phi_{\sigma}$ can be expressed as a tensor product of a single site states $\ket{+}$, $\ket{-}$ and $\ket{0}$.
If the first spin-$1/2$ of the chain is in the $\ket{\uparrow}$ state, then for the first site we cannot have a $\ket{-}$ state but only $\ket{+}$ or $\ket{0}$.
In the latter case, we still must have the first non-zero character to be a $+$ in $\sigma$ in order to satisfy the construction of the valence bond state.
It can be verified that there has to be the same number of $+$’s and $-$’s alternating all along the $\sigma$ string, with no further restrictions on the number of $0$’s between them.

Therefore, a typical allowed state $\Phi_{\sigma}$ in the AKLT model could look like this:
\begin{equation}
    \Phi_{\sigma} = \ket{000+-0+-+0-+0-+-0}
    \label{eq:phi_string}
\end{equation}
A look at \eqref{eq:phi_string} reveals that is a sort of Néel order (antiferromagnetic order) if we ignore the $0$’s.
Still, we cannot predict what two spins in two distant sites will be, as we have no control on the number of the $0$’s.
Indeed, there is no \emph{local} order parameter that can be found to be non-zero in the Haldane phase and that can be used to distinguish this phase from the others. But, there is actually a \emph{non-local} order parameter, the string order parameter, that is able to reveal the hidden order of the Haldane phase.

In order to see how we can arrive at its definition, let us introduce the non-local unitary transformation
\begin{equation}
    U = \prod_{k=1}^{N} \prod_{j=2}^{k-1} \exp \bigg( i\pi S^z_j S^x_k \bigg),
\end{equation}
where $N$ is the number of sites, such that
Consider a typical AKLT state $\Phi_{\sigma}$, for example \eqref{eq:phi_string}.
On this state, the operator $U$ acts as
\begin{equation}
    U\Phi_{\sigma} = (-1)^{z(\sigma)} \Phi_{\bar{\sigma}},
\end{equation}
where $z(\sigma)$ is the number of $0$ characters in odd sites and $\bar{\sigma}$ is the new transformed string.
It is defined as follows:
\begin{itemize}
    \item if $\sigma_i = +$ (or $-$) and the number of non-zero characters to the left of the site $i$ is odd, then $\bar{\sigma}_i = -$ (or $+$).
    \item otherwise, $\sigma_i = \bar{\sigma}_i$
\end{itemize}
where $\sigma_i$ is the $i$-th character of the string $\sigma$.
In particular, if we apply this transformation on the allowed state \eqref{eq:phi_string}, it becomes:
\begin{equation}
    U\Phi_{\sigma} = \ket{000++0+++0++0+++0}.
\end{equation}
Then this unitary transformation aligns all the non-zero spins i.e.~if the first non-zero character is $+$ (or $-$) all the other non-zero characters become $+$ (or $-$).
It is also evident that $U^{-1} = U$.

Under the action of $U$, the spin operators transform as follows:
\begin{equation}
    \begin{split}
        \widetilde{S}^x_j & = US_j^x U^{\dagger} = S_j^x \left(e^{i\pi \sum_{l>j} S_l^x } \right), \\
        \widetilde{S}^y_j & = US_j^y U^{\dagger} = \left( e^{i\pi \sum_{l<j} S_l^z } \right) S_j^y \left( e^{i\pi \sum_{l>j} S_l^x } \right), \\
        \widetilde{S}^z_j & = US_j^z U^{\dagger} = \left( e^{i\pi \sum_{l<j} S_l^z } S_j^z \right).
    \end{split}
    \label{eq:newoperators}
\end{equation}
Notice that the local operators have been mapped onto non-local operators, as they contain a sum of spin operators acting on different sites.
This is not surprising, given that $U$ itself is a non-local unitary transformation.

It is reasonable to expect that also the local Hamiltonian $H$ is mapped onto a non-local one $\widetilde{H} = U H U^{-1}$, but it turns out that $\widetilde{H}$ is still, in fact, local:
\begin{equation}
    \widetilde{H} = J \sum_{j} \big[h_j + \beta (h_j)^2 \big],
\end{equation}
where
\begin{equation}
    h_j = -S^x_j S^x_{j+1} + S_j^y e^{i \pi \left(S_j^z + S^x_{j+1}\right)} S^y_{j+1} - S^z_j S^z_{j+1}
\end{equation}
The transformed Hamiltonian $\widetilde{H}$ still has the same symmetries of $H$, but they may not be local anymore.
Actually, the only local symmetry of $H$ is related to its invariance under rotations of $\pi$ about each coordinate axis.
This symmetry group is equivalent to $\mathbb{Z}_2\times \mathbb{Z}_2$: indeed, the product of two $\pi$-rotations about two different axes produce a $\pi$- rotation about the third one.

It is possible to prove \cite{affleck1988valence} that at the AKLT point the transformed Hamiltonian has four ground states, which are product states and break such symmetry.
These four degenerate ground states of $H_{\text{AKLT}}$ converge to a single ground state in the infinite volume limit.
The same is not true for the ground states of $\widetilde{H}_{\text{AKLT}}$, as they converge to four distinct states in the infinite volume limit, even though the two Hamiltonians are related by a unitary transformation.
In a sense, the non-locality of the transformation $U$ does not guarantee a one-to-one correspondence between the ground states in the infinite volume limit.

Finally, we can understand the role of the string order parameter \eqref{eq:newoperators}.
In fact, it is straightforward to verify that
\begin{equation}
    S^{\alpha}_1\bigg( \prod_{k=2}^{r-1} e^{i \pi S_k^{\alpha}}\bigg) S^{\alpha}_r  =-U^{-1}S_{1}^{\alpha}S_{r}^{\alpha} U.
\end{equation}
This shows that the NLOPs in \eqref{NLOP} reveal the ferromagnetic order in the language of the non-local spins \eqref{eq:newoperators}
 or, equivalently, the breaking of the hidden symmetry in the original system.
Such a symmetry breaking holds in the whole Haldane phase, not just the AKLT model.
Indeed, in the dimer phase the symmetry is completely unbroken and the string order parameter \eqref{NLOP} will vanish for every $\alpha$.

\newpage
\clearpage
\bibliography{biblio}

\end{document}